\documentclass[nofootinbib]{article} 
\usepackage[utf8]{inputenc}
\usepackage[T1]{fontenc}
\usepackage{geometry}
\usepackage{amsmath,amssymb,amsfonts,bm}
\usepackage{graphicx} 
\usepackage{braket} 
\usepackage{authblk} 
\usepackage{lettrine}
\usepackage{color}
\usepackage{times}
\usepackage{hhline}
\usepackage{tabularx}
\usepackage{calc}
\usepackage{soul} 
\usepackage{mathbbol}
\usepackage{makecell}
\usepackage{subcaption}
\usepackage{multirow}
\usepackage{wrapfig}
\usepackage{xcolor} 
\usepackage[table]{xcolor}
\usepackage[normalem]{ulem}
\usepackage{booktabs}
\usepackage{parskip}
\usepackage{siunitx}
\usepackage{pbox,cellspace}
\usepackage{colortbl}
\usepackage{setspace}
\usepackage{comment}
\usepackage{lineno}
\usepackage[numbers,sort&compress]{natbib}
\usepackage[colorlinks=true,linkcolor=blue,citecolor=blue,urlcolor=blue]{hyperref}
\renewcommand{\thefootnote}{\fnsymbol{footnote}}
\title{Local and itinerant orbital contributions to the gyromagnetic response of paramagnetic metals}
\author[1,2,*]{Garima Ahuja}
\author[2,3,4]{Mirco Sastges}
\author[5]{Dongwook Go}
\author[1,6]{Shobhana Narasimhan}
\author[2,4,$\dagger$]{Stefan Bl\"ugel}
\author[2,3,$\ddagger$]{Yuriy Mokrousov}
 
\affil[1]{Theoretical Sciences Unit and School of Advanced Materials, Jawaharlal Nehru Centre for Advanced Scientific Research, Bengaluru, 560064 Karnataka, India}
\affil[2]{Peter Gr\"unberg Institute, Forschungszentrum J\"ulich and JARA, 52425 J\"ulich, Germany} 
\affil[3]{Institute of Physics, Johannes Gutenberg University Mainz, 55099
Mainz, Germany}
\affil[4]{Institute for Theoretical Physics, RWTH  Aachen University, 52056 Aachen, Germany}
\affil[5]{Department of Physics, Korea University, Seoul 02841, South Korea}
\affil[6]{International Centre for Theoretical Sciences, Shivakote, Hesaraghatta, Bengaluru, 560089, Karnataka, India} 

\date{}
\begin{document}
\maketitle 
\footnotetext[1]{\hyperlink{mailto:garima@jncasr.ac.in}{garima@jncasr.ac.in}}
\footnotetext[2]{\hyperlink{mailto:s.bluegel@fz-juelich.de}{s.bluegel@fz-juelich.de}}
\footnotetext[3]{\hyperlink{mailto:y.mokrousov@fz-juelich.de}{y.mokrousov@fz-juelich.de}}
\renewcommand{\thefootnote}{\arabic{footnote}}
\section*{Abstract}
Orbitronics exploits orbital angular momentum to generate and manipulate angular-momentum currents for controlling magnetisation, offering new approaches to magnetic memory and logic. A quantitative description of this process requires establishing how orbital angular momentum contributes to magnetic moments and, consequently, to the gyromagnetic response that governs magnetisation dynamics. Here, we present a first-principles framework for calculating the effective gyromagnetic tensor ($g^{\prime}$) of paramagnetic metals, consistently incorporating local and itinerant orbital contributions, spin-orbit coupling, and both state- and occupation-change effects. Applying the framework to V and Pt, we find that orbital paramagnetic responses are comparable to or larger than spin contributions and substantially modify $g^{\prime}$ from the free-electron value. The atom-centered approximation can underestimate the orbital response by approximately a factor of two, while neglecting occupation changes produces significant errors in the calculated $g^{\prime}$. Our calculated $g^{\prime}$ values for V and Pt agree with experiment, while their pronounced dependence on band filling reveals a route to engineering angular-momentum responses through electronic-structure control. The results establish the modern theory of orbital magnetism as an essential ingredient for describing gyromagnetic response and provide a framework for identifying materials with enhanced orbital functionalities.

\section*{Introduction} 
In crystalline solids, external perturbations can induce orbital responses that are comparable to, or even exceed, their spin counterparts. Even when the ground-state orbital angular momentum (OAM) is quenched by the crystal field, electronic states can acquire substantial OAM in response to external  fields~\cite{go2018,jo2018,gold_xmcd}. Recent work has predominantly focused on electric-field-induced orbital phenomena, including the orbital Hall effect (OHE) and orbital Edelstein effect (OEE)~\cite{kontani2009,go2018,jo2018,bhowal2020,choi2023,hayashi2023}. These advances form the basis of orbitronics, which seeks to generate, transport, and manipulate OAM for controlling magnetic states and developing low-power memory and logic devices~\cite{go2021,wang2025,jo2024_Rev.}.

Equally fundamental, but much less explored, is the response of orbital angular momentum to a magnetic field. This response contributes to both the induced magnetic moment and the total angular momentum, whose relation is characterised by the "gyromagnetic" tensor $g^{\prime}$. The gyromagnetic response sets the scale of magnetic precession and, together with damping, influences the timescale of relaxation. It also reveals the relative orbital and spin character of the induced response: when mixed spin--orbital contributions are negligible, the orbital- and spin-dominated limits approach $g^{\prime}=1$ and $g^{\prime}=2$, respectively. Understanding the microscopic origin of $g^{\prime}$ is therefore important for magnetic resonance and magnetisation dynamics, as well as for orbitronics, where generated or transported orbital angular momentum can ultimately couple to magnetic degrees of freedom. Despite this importance, the microscopic orbital contributions to $g^{\prime}$ in crystalline metals remain poorly understood. In particular, it is unclear how much of the gyromagnetic response originates from local orbital circulation and how much from itinerant orbital motion, and how these compete with spin and mixed spin-orbital contributions. 

The importance of orbital contributions to the magnetic response of metals was recognised early by Kubo and Obata, who pointed out that conventionally computed values of paramagnetic susceptibility $\chi$ did not agree with experiment presumably
due to the neglect of orbital effects~\cite{kubo1956}. Subsequent semi-empirical and first-principles band-structure approaches incorporated orbital contributions using atom-centered or conventional decompositions~\cite{mori1969,mori1968,yasui1991}. Such treatments, however, cannot fully capture the itinerant nature of orbital magnetism in a periodic crystal.
This limitation becomes particularly consequential for the gyromagnetic tensor $g^{\prime}$, which relates the total magnetic moment to the total angular momentum. Although closely related to the spectroscopic splitting tensor $g$, the two quantities do not generally coincide~\cite{scott1962,kittel1949}. Previous \textit{ab initio} studies have primarily addressed the spectroscopic $g$-tensor~\cite{cehovin2004,ceresoli2010,pickard_mauri2002,schreckenbach}, whereas a first-principles description that resolves the local and itinerant orbital contributions to the gyromagnetic tensor $g^{\prime}$ of metallic systems remains lacking.
Previous theoretical studies have established general formulations of orbital magnetic susceptibility and elucidated its geometric origins~\cite{gao,raoux2015}. Here, we focus on the paramagnetic contributions to magnetic susceptibility in metals and $g^{\prime}$, and establish microscopically how spin, orbital, and mixed spin--orbital contributions combine to determine the gyromagnetic response.   
 
The orbital angular momentum in a periodic crystal cannot, in general, be represented as a sum of atom-centered contributions. In particular, the position operator entering $\hat{\mathbf{L}}=\hat{\mathbf{r}}\times\hat{\mathbf{p}}$ is not well defined in the Bloch representation, while the physically observable orbital magnetisation contains both local and itinerant contributions. Consequently, calculating orbital contributions to $\chi$ and $g^{\prime}$ from first principles presents a significant challenge due to the complexity in treating the OAM operator in terms of Bloch states, the treatment of which has progressed only very recently~\cite{lee2026,sastges2026,lopez2012}. In this work, we address the challenge by employing a rigorous, gauge-covariant formulation of the OAM operator within a Wannier representation, allowing local and itinerant orbital contributions to be treated on the same footing.

In the following, we present an approach to computing paramagnetic susceptibility $\chi$, and thus $g^{\prime}$, using perturbation theory in the linear-response framework. Our expressions contain all the terms in the Kubo-Obata formula~\cite{kubo1956}, as well as additional terms capturing contributions from changes in the occupations of the electronic states – an important contribution in metals. We apply the framework to bcc V and fcc Pt as prototype paramagnetic metals representing weak and strong spin-orbit-coupling (SOC) regimes, as well as weak and strong Stoner susceptibility regimes, respectively. We find that the orbital paramagnetic 
susceptibility can be very large in transition metals, and the contributions from both the ‘occupation-change’ and the ‘state-change’ terms play a crucial role in modifying $g'$ from the free-electron value. We further demonstrate that the atom-centered approximation (ACA) to the orbital moment can substantially underestimate the orbital susceptibility and the use of Berry phase theory, commonly known as the modern theory of orbital magnetism (OM)~\cite{thonhauser2005,xiao2005,shi2007}, is crucial for providing precise material-specific predictions. The framework enables the screening and design of materials based on their orbital, spin–orbital, and spin responses, thereby facilitating the identification of promising candidates for orbital- and spin-angular-momentum-based devices.

\section*{Results} 
\subsection*{Formalism}
In the Wannier representation, the modern theory of orbital magnetism includes contributions to OM from local circulation (LC) and itinerant circulation (IC)~\cite{thonhauser2005,lopez2012}. The LC contribution describes the self-rotation of Wannier functions about their centers and therefore represents a local bulk contribution. The IC contribution, on the other hand, originates from the circulation of Wannier-function centers; in a finite sample, this circulation is manifested as boundary currents. We include these contributions to OAM by calculating matrix elements of the OAM operator $\hat{\mathbf{L}}$ using the modern theory gauge-covariant formulation~\cite{lee2026,sastges2026}. The expression for the general matrix elements of the OAM operator at any $\mathbf{k}$-point can be written as:   
\begin{equation} \label{angularm}
{\textbf{{L}}}_{nm}= \langle  u_{n\textbf{k}} \vert \hat{\textbf{L}} \vert u_{m\textbf{k}} \rangle = 
 \frac{ie}{2\mu_\mathrm{B}} \langle \partial_{\textbf{k}} u_{n\textbf{k}} \vert \times \left(  \hat{H}_{\textbf{k}} + \frac{E_{n\textbf{k}} + E_{m\textbf{k}} }{2} - 2 E_\mathrm{F} \right) \vert \partial_{\textbf{k}}u_{m\textbf{k}} \rangle,
\end{equation}  
where $E_{n\mathbf{k}}$ represents the energy associated with the eigenstate $u_{n\mathbf{k}}$ of the cell-periodic Bloch Hamiltonian $\hat{{H}}_{\mathbf{k}}$. $E_\mathrm{F}$ is the Fermi energy, $\mu_\mathrm{B}$ is the Bohr magneton and $e$ is the positive electronic charge. 
The numerical implementation of the above expression involves evaluation of Berry-type matrices and the derivatives $\partial_{\mathbf{k}}$ are replaced with occupation-weighted covariant derivatives (details can be found in Section S2.1 of the Supplementary Information) to ensure covariance of the OAM operator under any gauge-transformation~\cite{lee2026,sastges2026}. In this formulation, the OAM operator correctly produces the OM in the ground-state in addition to including the local and itinerant orbital contributions in describing the off-diagonal elements of $\hat{\mathbf{L}}$.  
 
We define the components of the $g^{\prime}$-tensor~\cite{kittel1949,trukhanova2026}, $g^{\prime}_{\alpha\beta}$, as: $\mathbf{m}_{\alpha} = (\mu_\mathrm{B}/\hbar) g^{\prime}_{\alpha \beta}\mathbf{J}_{\beta}$, where $\mathbf{m}_{\alpha}$ is the total magnetic moment,  $\mathbf{J_{\beta}}$ is the total angular momentum, $\alpha$ and $\beta$ are Cartesian directions, and $\hbar$ is the reduced Planck's constant.
The interaction of the orbital and spin angular momenta, represented by the $\hat{\mathbf{L}}$ and $\hat{\mathbf{S}}$ operators, respectively, with an external magnetic field B is described by the Zeeman Hamiltonian:
\begin{equation}\label{eq:zeeman_ham}
\hat{H}_\mathrm{Z} = \frac{\mu_\mathrm{B}}{\hbar} \mathbf{B}\cdot(g^{0}_\mathrm{L}\hat{\mathbf{L}}+g^{0}_\mathrm{S}\hat{\mathbf{S}}),
\end{equation}   
where $g^{0}_\mathrm{L}$ and $g^{0}_\mathrm{S}$ denote the orbital and spin $g$-factors for a free electron, respectively, with $g^{0}_\mathrm{L} = 1$ and $g^{0}_\mathrm{S} \approx 2$. Treating ${H_\mathrm{Z}}$ as a perturbation, the changes in values of the orbital and spin moments can be determined as~\cite{vanderbilt2018,go2025orb_pumping}:  
\begin{equation}\label{eq:del_l}   
\delta \langle L_{\alpha}\rangle
= (\chi^\mathrm{LL}_{\alpha \beta} + \chi^\mathrm{LS}_{\alpha \beta})  B_{\beta},\quad
\delta \langle 2S_{\alpha}\rangle
= (\chi^\mathrm{SS}_{\alpha \beta} + \chi^\mathrm{SL}_{\alpha \beta}) B_{\beta}.
\end{equation}  
$\chi^\mathrm{LL}$ and $\chi^\mathrm{SS}$ are the orbital and spin zero-field linear susceptibility tensors, respectively, describing pure orbital and pure spin responses to a magnetic field at zero field. 
We combine the mixed spin and orbital responses $\chi^\mathrm{LS}$ and $\chi^\mathrm{SL}$ to define the total spin-orbital (SO) susceptibility: $\chi^\mathrm{SO}$ = $\chi^\mathrm{LS}$ + $\chi^\mathrm{SL}$. 

We distinguish two distinct contributions to each of the three susceptibilities: 
the first arises from the field-induced modification of the electronic states themselves  ($\chi^{\rm{state\text{-}change}}_{\alpha \beta}$), while the second originates from the redistribution of their occupations  ($\chi^{\mathrm{occ\text{-}change}}_{\alpha \beta}$) [The derivation is given in Sections S1.2 and S1.3 of the Supplementary Information]: 
\begin{equation}
   \chi_{\alpha \beta} = \chi^{\rm{state\text{-}change}}_{\alpha \beta} + \chi^{\mathrm{occ\text{-}change}}_{\alpha \beta}. 
\end{equation}
The state-change contribution to, for example $\chi^\mathrm{LL}_{\alpha \beta}$, is given by the Kubo expression:
\begin{equation}\label{eq:suscep_ll_state}
\begin{split} \chi^{\mathrm{LL},\mathrm{state\text{-}change}}_{\alpha \beta} = & \sum_n \chi^{\mathrm{LL},\mathrm{state\text{-}change}}_{\alpha \beta,n}=\frac{\mu_0 \mu_\mathrm{B}^2}{N_k V \hbar^2}
\sum_{\textbf{k}} \sum_{n,m} (f_{m\textbf{k}} - f_{n\textbf{k}}) \\ & \times \frac{\langle u_{n\textbf{k}} | L_\alpha | u_{m\textbf{k}} \rangle \langle u_{m\textbf{k}} | L_\beta |   u_{n\textbf{k}} \rangle}{E_{n\textbf{k}} - E_{m\textbf{k}} + i \Gamma},
\end{split}   
\end{equation} where the use of gauge-covariant formalism for the OAM
operator ensures there is no ambiguity when obtaining terms of the form $\langle u_{n\textbf{k}} | L_\alpha | u_{m\textbf{k}} \rangle$.
In the expression, $V$ is the volume of the unit cell, the term $(f_{m\mathbf{k}} - f_{n\mathbf{k}})$ denotes the difference in occupations (given by the Fermi-Dirac distribution function) of the Bloch states represented by  $m\mathbf{k}$ and $n\mathbf{k}$, $\mu_{0}$ is the permeability of free space, $\mu_\mathrm{B}$ is the Bohr magneton, and $\Gamma$ is an infinitesimally small real number. Note that the equations for susceptibilities are in SI units where $\chi$ is dimensionless.

\begin{figure}
    \centering     \includegraphics[width=0.33\linewidth]{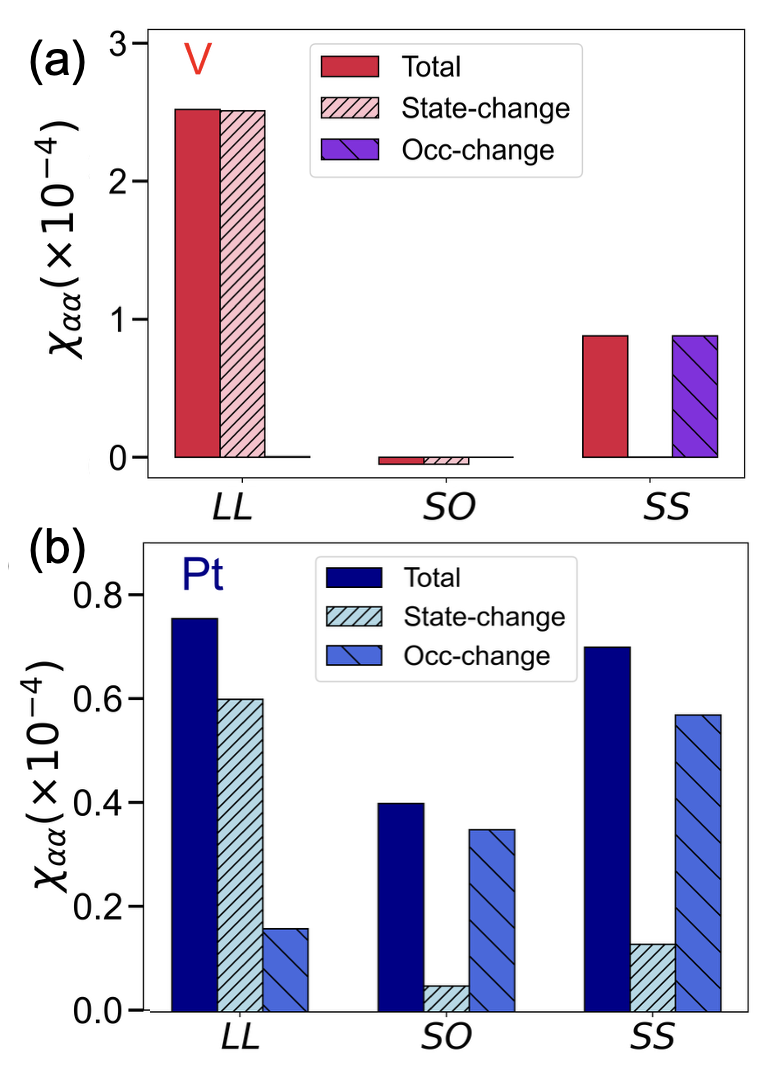}
    \caption{Orbital (LL), spin-orbital (SO), and spin (SS) paramagnetic susceptibilities. Decomposition into total, state-change and occupation-change terms, calculated for (a) V and (b) Pt. The values are in SI units. Note that the $y$-axis scales  are different in (a) and (b).}
    \label{fig:num_values_chi}
\end{figure}
The second contribution to susceptibilities is driven by the changes in occupation of the energy state as the magnetic field is applied to the system. These are given by, for example, 
\begin{equation}\label{eq:ll_occ_suscep}  
\begin{split}
 \chi^{\mathrm{LL}, \mathrm{occ\text{-}change}}_{\alpha \beta} = &
\sum_n \chi^{\mathrm{LL}, \mathrm{occ\text{-}change}}_{\alpha \beta,n}=-
\frac{\mu_0 \mu_\mathrm{B}^2}{N_k V \hbar^2}
\sum_{\mathbf{k}}\sum_{n}
f^\prime(E_{n\mathbf{k}})
\\ & \times \langle u_{n\mathbf{k}}|L_{\alpha}|u_{n\mathbf{k}}\rangle \langle u_{n\mathbf{k}}| L_{\beta}|u_{n\mathbf{k}}\rangle,
\end{split}
\end{equation}
where $f^\prime(E_{n\mathbf{k}})=\left.\frac{df(E_{n\mathbf{k}})}{dE}\right|_{E=E_{n\mathbf{k}}}$ is the derivative of the Fermi-Dirac distribution function at $E_{n\mathbf{k}}$. 

The expressions for $\chi^{\mathrm{SO}}_{\alpha \beta}$ and $\chi^{\mathrm{SS}}_{\alpha \beta}$ for both state-change and occupation-change terms can be defined by the appropriate replacement of orbital angular momentum operators $\hat{\mathbf{L}}$ with spin operators 2$\hat{\mathbf{S}}$. The equations are given in Sections S1.2 and S1.3 of the Supplementary Information. These expressions are evaluated using first-principles Wannier interpolation calculations (computational details are given in the Methods Section). 
In a cubic paramagnetic crystal, whose equilibrium state preserves the crystal symmetry, the
uniform, zero external magnetic field linear-response susceptibilities 
$\chi$ are isotropic tensors of rank 2  satisfying
\begin{equation}
\chi^\mathrm{AB}_{\alpha\beta}=\chi^\mathrm{AB}\delta_{\alpha\beta},
\qquad \mathrm{A,B}\in\{\mathrm{L,S}\}.
\end{equation}
Consequently, the gyromagnetic tensor constructed from the magnetic-moment and angular-momentum responses is isotropic,
\begin{equation}
g^\prime_{\alpha\beta}=g^\prime \delta_{\alpha\beta}.
\end{equation} $g^{\prime}$ can be constructed using the knowledge of the susceptibilities (see Section S1.4 of Supplementary Information for a derivation):     
\begin{equation}\label{eq:g}
g^{\prime} = \frac{(\chi^\mathrm{LL} +  \chi^\mathrm{SO} +  \chi^\mathrm{SS})}{(\chi^\mathrm{LL} + \frac{3}{4}\chi^\mathrm{SO} + \frac{1}{2}\chi^\mathrm{SS})}.
\end{equation} 

External fields produce field-dependent quartic terms whose spin-spin and spin-to-orbital contributions proved very small and are not further considered.
\subsection*{Materials selection}
We consider V and Pt as two prototype paramagnetic metals that provide complementary regimes of SOC, orbital response and interacting spin enhancement. V is an early 3d transition metal representing a weak-SOC regime in which orbital currents could emerge without strong relativistic coupling. Recent experimental and theoretical studies have identified V as an efficient source of orbital currents~\cite{jo2018,vijayan2025}. Pt, in contrast, is a late 5d transition metal with strong SOC, providing a complementary regime in which orbital and spin degrees of freedom are strongly coupled. The competition between kinetic energy and electron-electron interactions gives rise to a strong Stoner enhancement of the spin susceptibility. Beyond its established role in spin Hall physics and spin-orbit torque applications, Pt is also a prominent orbitronics material owing to its strong orbital Hall response and efficient generation and transport of orbital angular momentum~\cite{tanaka2008,Pt_sala}. The coexistence of orbital and spin Hall effects further enables the conversion of orbital currents into spin currents and torques, linking orbital transport to magnetisation control. Studying V and Pt allows us to examine whether the differences between the ACA and the modern theory of orbital magnetism are specific to a particular electronic regime or persist across qualitatively different paramagnetic metals.

\subsection*{Paramagnetic susceptibility and g$^{\prime}$-tensor}\label{ssec:susc}
\begin{figure}[t]
    \centering   \includegraphics[width=0.4\linewidth]{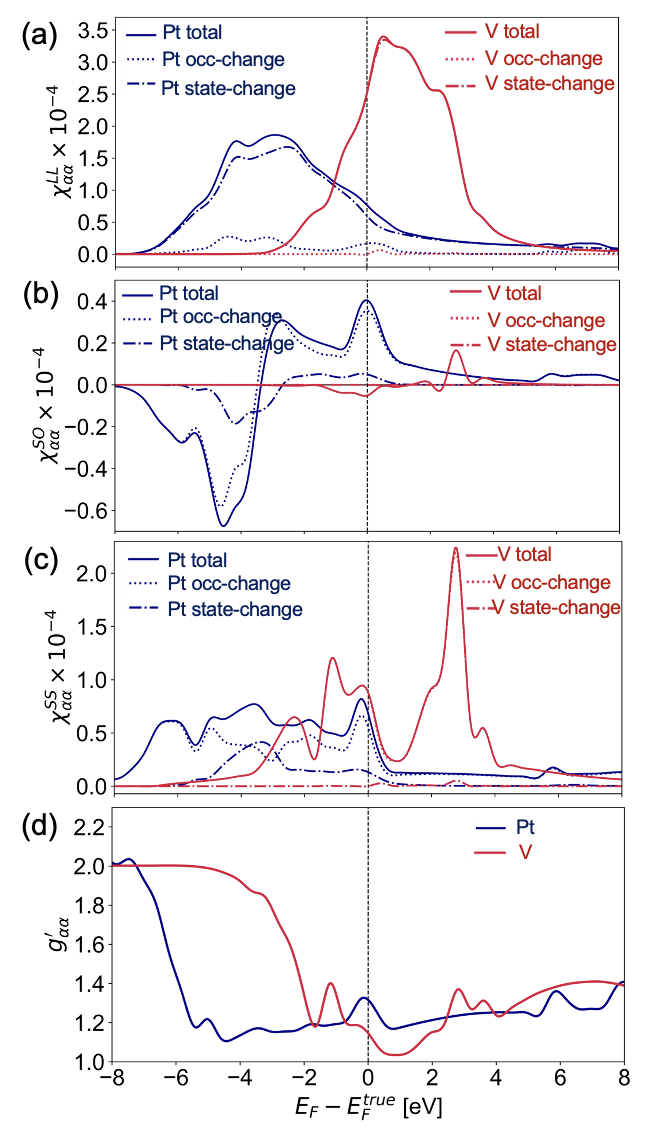}    \caption{Band-filling dependence of paramagnetic susceptibilities and the gyromagnetic tensor $g^{\prime}$. The susceptibilities are decomposed into state-change and occupation-change contributions. Red and dark blue curves represent V and Pt, respectively. (a) Orbital susceptibilities, (b) spin-orbital susceptibilities, (c) spin susceptibilities, and (d) gyromagnetic tensor $g^{\prime}$. E$_{\mathrm{F}}^{\mathrm{true}}$ denotes the Fermi energy of the pristine system, in the absence of defects or impurities.} \label{fig:band_filling}
\end{figure} 
Figure~\ref{fig:num_values_chi} summarises the calculated $zz$ components of the orbital, spin, and mixed spin--orbital susceptibilities, $\chi^\mathrm{LL}$, $\chi^\mathrm{SS}$ and $\chi^\mathrm{SO}$, respectively. A striking result is the orbital dominance of the magnetic response in V: $\chi^\mathrm{LL}=2.52\times10^{-4}$ is nearly three times the spin susceptibility, $\chi^\mathrm{SS}=0.88\times10^{-4}$, whereas the mixed contribution, $\chi^\mathrm{SO}=-0.05\times10^{-4}$, is small and opposes the positive orbital and spin responses. Thus, weak SOC does not preclude a substantial orbital magnetic response, consistent with previous reports of strong orbital effects in V~\cite{vijayan2025,jo2018}.

Pt exhibits a different balance. Its orbital and spin susceptibilities, $\chi^\mathrm{LL}=0.76\times10^{-4}$ and $\chi^\mathrm{SS}=0.70\times10^{-4}$, are comparable, while the mixed response, $\chi^\mathrm{SO}=0.40\times10^{-4}$, contributes approximately $21\%$ of the total. The substantial mixed contribution is consistent with stronger spin-orbital coupling in Pt. These contrasting results show that orbital magnetism is essential even in a weak-SOC metal, while mixed spin-orbital contributions become quantitatively important in Pt. A description based on the spin susceptibility alone therefore cannot capture the gyromagnetic response of either material. 
The analytical expressions for the non-interacting spin susceptibility $\chi^\mathrm{SS}$ and mixed spin-to-orbital susceptibility $\chi^\mathrm{LS}$, derived in Section~S1 of the Supplementary Information, are tested independently using finite-field density functional theory calculations. We apply a uniform spin-Zeeman field throughout the unit cell and calculate the induced spin and orbital magnetic moments, $\delta \mathbf{m}_\mathrm{S}$ and $\delta \mathbf{m}_\mathrm{L}$, while keeping the ground-state Kohn-Sham potential fixed. The susceptibilities are extracted from the corresponding moment-field slopes in the limit $B\rightarrow0$ (see Section~S3.1 of the Supplementary Information). For V and Pt, respectively, the finite-field spin susceptibilities are $0.80\times10^{-4}$ and $0.726\times10^{-4}$, compared with the Kubo values of $0.88\times10^{-4}$ and $0.70\times10^{-4}$, corresponding to differences of approximately $9$~\% and $3.7$~\%. Within the ACA, the finite-field mixed susceptibility of Pt is $0.115\times10^{-4}$, within $4.5$\% of the Kubo value of $0.11\times10^{-4}$. For V, both approaches yield a nearly vanishing mixed response: $-0.0015\times10^{-4}$ from finite-field calculations and approximately zero from the Kubo expression. These comparisons provide an independent numerical check of the spin and ACA-based mixed susceptibilities.
\begin{table}[b!]  
    \centering    
    \renewcommand{\arraystretch}{1.3}
    \begin{tabular}{c c c c c c c}\hline
         Material& $g^\prime$ (total)&$g^\prime$(state-change)&$g^\prime$(ACA)&$g^\prime$(total, int) & $g^\prime$(ACA, int) & $g^\prime$(expt.)\\  \hline  
         V& 1.15& 1.00&1.20& 1.32 & 1.40&1.18 $\pm$ 0.10~\cite{huguenin_expt},
         \\& & & & & &1.20 $\pm$ 0.10~\cite{gyro_v}\\
         Pt& 1.32& 1.11&1.43& 1.61 & 1.73& 1.62 $\pm$ 0.08 ~\cite{huguenin_expt}\\\hline \end{tabular}
    \caption{Gyromagnetic tensor  $g^\prime$ (isotropic value) for V and Pt. The columns with numerical values are obtained using (i) the modern theory approach to orbital angular momentum, (ii) modern theory including only state-change contributions, (iii) the atom-centered approximation (ACA), (iv) modern theory with interacting susceptibilities, (v) ACA with interacting susceptibilities, and (vi) experiment~\cite{huguenin_expt,gyro_v}.}
    \label{tab:g}
\end{table}

To identify the microscopic origin of these responses, we separate each susceptibility into state-change and occupation-change contributions. In V, $\chi^\mathrm{LL}$ and $\chi^\mathrm{SO}$ are dominated by interband state changes, whereas the spin susceptibility $\chi^\mathrm{SS}$ is dominated by the Fermi-surface occupation-change contribution. The occupation-change term in $\chi^\mathrm{SO}$ is negative, indicating an opposing mixed spin--orbital contribution. In Pt, occupation changes dominate both $\chi^\mathrm{SO}$ and $\chi^\mathrm{SS}$, demonstrating their importance for the magnetic response of metals.
 
The susceptibilities also depend strongly on band filling [Fig.~\ref{fig:band_filling} (a-c)]. The orbital and spin responses remain positive throughout the investigated energy range, whereas the mixed susceptibility can change sign. This sign change follows  qualitatively Hund's third rule where the spin and orbital angular momentum couple opposite (parallel) for less (more) than half  band filling, reminiscent of the filling-dependent spin--orbit correlations discussed in Ref.~\cite{lee_ohe}, although the two quantities describe different aspects of the electronic response. In both metals, $\chi^\mathrm{LL}$ varies smoothly and is predominantly governed by state changes. In Pt, $\chi^\mathrm{SS}$ remains approximately constant as band filling decreases, despite a redistribution of its weight from state-change to occupation-change contributions. Thus, similar total susceptibilities can conceal substantially different microscopic origins. The filling dependence suggests opportunities for tuning the balance of orbital, spin, and mixed responses, and consequently $g^{\prime}$, through electronic-structure control.
 
Table~\ref{tab:g} presents the calculated values of $g^{\prime}$ for V and Pt. The values $g^{\prime}=1.15$ and $1.32$, respectively, are substantially below the free-electron value $g_{\mathrm e}\simeq 2.0023$. For V, the calculated value agrees well with the experimental result of $1.18\pm0.10$, whereas for Pt it underestimates the measured value of $1.62\pm0.08$~\cite{huguenin_expt}. We discuss the role of exchange--correlation enhancement in this discrepancy below. To our knowledge, no previous theoretical calculations of $g^{\prime}$ have been reported for either metal. Including only state-change contributions yields $g^{\prime}=1.00$ for V and $1.11$ for Pt, increasing the discrepancies with experiment in both cases. Occupation changes therefore make a substantial contribution to the gyromagnetic response and are essential for its quantitative description in these metals.

\begin{figure}[t]
    \centering    \includegraphics[width=\linewidth]{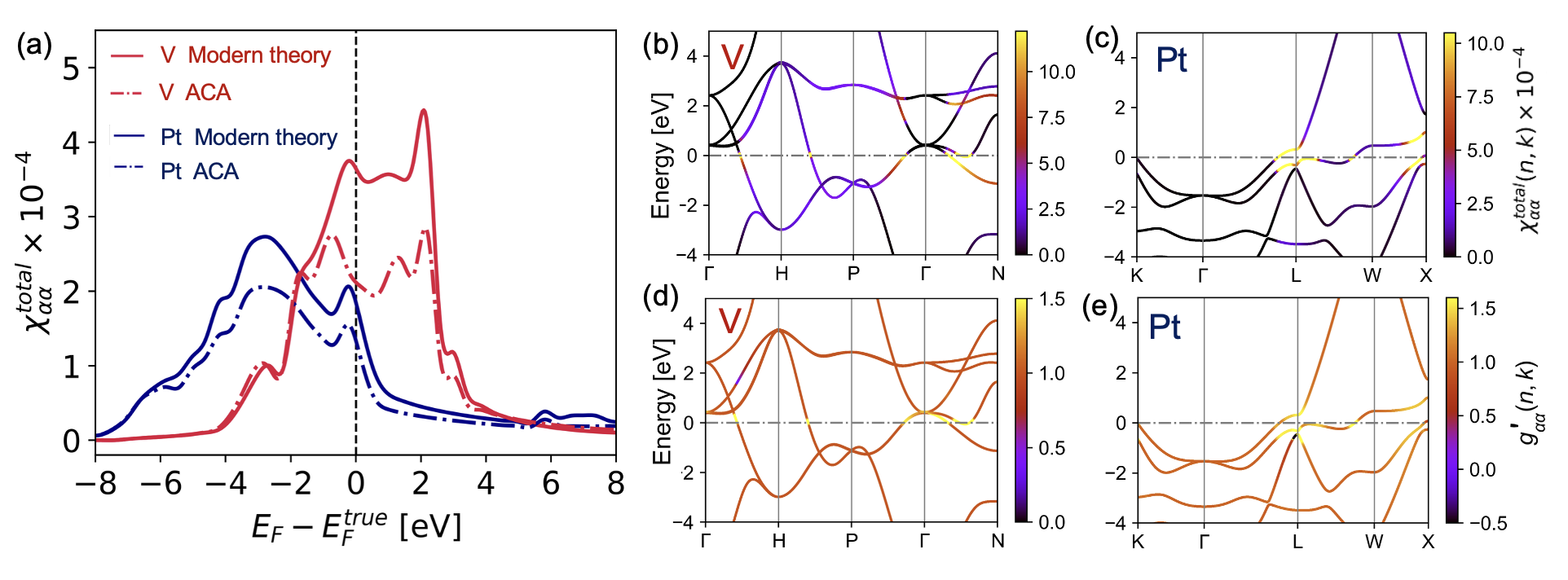}
    \caption{Energy-, band-, and $\mathbf{k}$-resolved paramagnetic susceptibilities and gyromagnetic tensor $g'$ from modern theory. (a) Comparison of the total paramagnetic susceptibility $\chi$ obtained using modern theory and the atom-centred approximation (ACA) for V (red) and Pt (dark blue). (b,c) Band- and $\mathbf{k}$-resolved contributions to the total susceptibility, (d,e) band- and $\mathbf{k}$-resolved $g^{\prime}$. The colour scales indicate the magnitude and sign of the corresponding contributions.}
    \label{fig:modern_aca}
\end{figure}

Figure \ref{fig:modern_aca} (a) shows the comparison of values of $\chi$ obtained using either the modern theory (solid lines) or the ACA (dot-dashed lines) when computing $\mathbf{L}$. ACA fails to capture the orbital response quantitatively. The ACA, while being  easier to implement, has the disadvantage that it does not take into account non-local and itinerant contributions to the OAM, and the results are sensitive to the choice of the basis set~\cite{lee2026,hanke2016}. In Section S2 of the Supplementary Information we further show that for both V and Pt, the value of $\chi^\mathrm{LL}$ in particular is severely underestimated by the ACA, being smaller by a factor of $\sim$2 when compared to the values obtained using the modern theory. We also present an analysis of the source of this discrepancy, showing that it cannot be attributed solely to the disregard of itinerant circulation terms in the ACA. The contribution to OAM captured by the ACA does not directly correspond to the local circulation term in the modern theory. This is because the LC part of the OAM operator also contains non-local terms involving matrix elements among Wannier functions centered
on different atoms. We therefore emphasise that it is vital to use the modern theory to obtain accurate results. 
  
Next, we examine how the choice between the modern theory and the ACA affects the values of $g^{\prime}$. Table~\ref{tab:g} lists the values of \(g^{\prime}\) obtained within the ACA. Since the orbital effects are stronger in V than in Pt, the value of $g^{\prime}$ deviates more from 2. This effect is not fully captured by the use of ACA for V. At first sight an examination of Table~\ref{tab:g} might suggest that the ACA does not introduce as large an error in the value of $g^{\prime}$ for Pt; however this is due to a fortuitous cancellation of errors in evaluating the numerator and the denominator in Eq.~(\ref{eq:g}). 
Further, we see that $g^{\prime}$ for both metals shows a non-trivial dependence on band filling especially near the true Fermi energy of the system, see Figure~\ref{fig:band_filling}  (d), implying that the magnitude can be adjusted by fine tuning the position of the Fermi level. In particular for the case of V we observe a clear transition from purely spin-dominated ($g^{\prime}=2$) to purely orbital-dominated ($g^{\prime}=1$) values of $g^{\prime}$ with band filling, as at low energies the $s$-state susceptibility is largely spin-dominated, while  $\chi^\mathrm{LL}$ is largest $-$ exceeding other components by an order of magnitude $-$ around the true Fermi energy.    

For Pt, the sizeable discrepancy between the bare-response and experimental $g^{\prime}$ values in Table~\ref{tab:g} motivates including exchange-correlation feedback, particularly given its strongly Stoner-enhanced spin susceptibility. We therefore extend the fixed-potential finite-field calculations to self-consistent calculations, allowing the Kohn--Sham potential to respond to the applied spin-Zeeman field. Comparing the resulting susceptibilities with their non-interacting counterparts gives a spin enhancement factor $S_{\mathrm S}=3.2$, whereas the mixed spin-to-orbital response is strongly suppressed, with $S_{\mathrm{LS}}=0.0287$ (Supplementary Section~S3.2).

To estimate the accompanying orbital renormalisation, we use these response factors in a minimal effective spin--orbital interaction kernel with no direct orbital--orbital feedback term (Supplementary Section~S3.3). Applying this procedure to the modern-theory susceptibilities predicts a reduction of $\chi^\mathrm{LL}$ by approximately $7.5$~\% and yields $g^{\prime}_{\mathrm{int}}=1.61$, in agreement with the experimental value of $1.62\pm0.08$. The orbital correction arises indirectly through the coupled response channels. Transferring the ACA-based finite-field response factors to the modern-theory susceptibilities is an assumption of this estimate; a direct orbital exchange--correlation contribution is not evaluated here.

For V, the bare-response value $g^{\prime}=1.15$ already agrees well with the experimental result of $1.18\pm 0.10$. Including self-consistent feedback, with the calculated Stoner enhancement facts $S_{\mathrm S}=2.7$ and $S_{\mathrm{LS}}=3.69$, increases the calculated value to $g^{\prime}_{\mathrm{int}}=1.32$, slightly above the reported experimental intervals. Thus, exchange--correlation feedback substantially improves the gyromagnetic response of Pt, but does not improve the agreement for V. The mixed-response factor for V should also be interpreted cautiously because its bare mixed susceptibility is nearly vanishing.

The calculated total interacting paramagnetic susceptibilities nevertheless compare favourably with the measured total susceptibilities of both metals. For Pt, we obtain $2.95\times10^{-4}$, compared with $2.66\times10^{-4}$ at 290~K~\cite{CRC_handbook_ofmag} and $2.63\times10^{-4}$ at 293~K~\cite{Childs01101959}. For V, the calculated value of $4.71\times10^{-4}$ compares with $4.28\times10^{-4}$ at 290~K~\cite{CRC_handbook_ofmag} and $4.45\times10^{-4}$ at 293~K~\cite{Childs01101959}. To enable a direct comparison with our results, the susceptibility values reported in these references were converted to consistent units. The calculated values exceed experiment by approximately $11$--$12$~\% for Pt and $5$--$10$~\% for V; diamagnetic contributions are omitted from the calculations. This comparison also illustrates that agreement in the total susceptibility does not necessarily imply the same accuracy in $g^{\prime}$, which depends on the relative weights of the response channels.

Using the renormalised ACA susceptibilities instead gives $g^{\prime}=1.73$ for Pt and $1.40$ for V, compared with the modern-theory values of $1.61$ and $1.32$, respectively. The larger ACA values are consistent with its underestimated orbital response, which shifts the balance towards the spin-dominated limit. Currently the  exchange-correlation feedback is included only through the spin and spin-orbital channels. These differences suggest, however, that the direct exchange--correlation feedback in the orbital channel may be important. Its consistent treatment would require the inclusion of the magnetic vector potential and a corresponding extension of the exchange--correlation functional. Both are beyond the scope of the present work and are left for future investigation. 

To gain better insight into how the details of the electronic structure affect $\chi$ and $g^\prime$, we plot the electronic band structure for V and Pt in Fig.~\ref{fig:modern_aca} (b,c); in these plots the colour represents the ($n,\mathbf{k}$)-resolved values of $\chi$, where $n$ is the band index and $\mathbf{k}$ is the wave vector. In V, the susceptibility is primarily determined by contributions from wider regions in energy and $\mathbf{k}$-space, and realised by band edges (see, e.g., the N-point) as well as dispersive bands. In Pt, we observe that $\chi$ is determined largely by the  contributions around the Fermi surface, with additional contributions from the vicinity of W--X and the L-point.  These distributions have an impact on the behaviour of $g^\prime$, see Fig.~\ref{fig:modern_aca} (d,e), where we present the distribution of the $\mathbf{k}$- and state-resolved values of $g^{\prime}$, evaluated according to Eq.~\ref{eq:g} with corresponding $(n,\mathbf{k})$-values of susceptibilities (see Section S1.4 of Supplementary Information). Thus defined, $g^{\prime}$ represents the $g^{\prime}$-factor when only a perturbation of a specific Bloch state by a magnetic field is considered. We observe that in V [Fig.~\ref{fig:modern_aca} (d)], in addition to a narrow region around the Fermi energy with enhanced values, $g^\prime\approx 1$ is more or less uniformly distributed in energy and $\mathbf{k}$-space. In the case of Pt, the region of enhanced values around $E_\mathrm{F}$ is much wider in energy, and the variations of $g^\prime$
in $\mathbf{k}$-space, taking place on the background of overall magnitude of about 1.2--1.4, are much more pronounced, especially around the X- and L- points. Remarkably, we uncover that, owing to large spin-orbit effects in Pt, for some Bloch states the value of state-dependent $g^\prime$ can even reach negative values, see a narrow region close to L-point along $\Gamma$-L. The presence of negative values leads to a prediction that in a strongly spin-orbit coupled solid, specific states may favour response to an applied magnetic field which is starkly different from the overall Fermi sea dynamics. In the Supplementary Information we compare this with the ($n,\mathbf{k}$)-resolved values of $g^{\prime}$ calculated using the ACA where we obtain positive values for all the states in Pt.

\section*{Discussion}
In this work, we have developed a first-principles framework for calculating the orbital response and the gyromagnetic tensor $g^{\prime}$ of paramagnetic crystalline metals. By resolving local and itinerant orbital contributions, spin and mixed spin-orbital responses, the framework reveals how these distinct channels determine the relation between induced angular momentum and magnetic moment. Our results demonstrate the importance of treating itinerant orbital motion and field-induced occupation changes explicitly when describing the magnetic response of metals. The inclusion of exchange-correlation feedback is essential for establishing the connection with the experimentally observed gyromagnetic response of strongly exchange enhanced metals like Pt. While the exchange enhancement predominantly acts on the spin channel, the coupled spin-orbital feedback also modifies the orbital response, despite the absence of a direct orbital exchange-correlation term in our treatment.  
 
We conclude that no single susceptibility $\chi$  universally controls the gyromagnetic response $g^{\prime}$, instead, its value is governed by the relative weights of the orbital, spin, and mixed responses. The pronounced Fermi-energy dependence of these susceptibilities identifies band filling as a possible means of tuning this balance and thereby the gyromagnetic response. Shifting the Fermi energy modifies both the electronic states contributing to the occupation-change response and the occupied-unoccupied transitions entering the state-change response, affecting the orbital, spin, and mixed susceptibilities differently. Since $g^{\prime}$ is governed by their relative weights, its control requires redistributing these responses; a common enhancement of all channels leaves $g^{\prime}$ unchanged. Our rigid-band analysis suggests that the orbital response reaches its maximum at fillings near the middle of the 3d and 5d transition-metal series, around paramagnetic Cr and Re, respectively. Consequently $g'$ is expected to deviate most strongly from the free-electron value in these regions, reflecting the enhanced contribution of orbital magnetism. This is also seen in the calculated values of $g'$ using interacting responses obtained with the rigid-band approximation. The values of $g'$ for paramagnetic Cr, Mn, Fe are 1.05, 1.30, 1.70, respectively, and for Re, Os, Ir are 1.32, 1.35, 1.37, respectively. Within the 3d series, the Stoner parameter is expected to vary only weakly across the series; we therefore approximate $I_{\mathrm{SS}}(Z)\simeq I_{\mathrm{SS}}(\mathrm{V})$ for the 3d series and  $I_{\mathrm{SS}}(Z)\simeq I_{\mathrm{SS}}(\mathrm{Pt})$ for the 5d series with $Z$ representing the chemical element. 

Within the rigid-band approximation, the calculated filling dependence suggests  carrier doping or alloying as possible routes to tailoring the gyromagnetic response. The observed possibility of tunability of the values of $\chi$ and $g'$ by tuning the position of the Fermi energy suggests a route to engineering damping, relaxation, and resonance properties relevant to spintronic and orbitronic devices and ferromagnetic resonance (FMR)-based spin pumping experiments. Establishing which experimental control parameter is most effective will require accounting for the accompanying changes in electronic structure and self-consistent response beyond the rigid-band approximation.  

The framework also provides a basis for investigating the dependence of the $g^{\prime}$ tensor on the strain as well as investigating lower-symmetry materials, where the anisotropy of $g^{\prime}$ can reveal an orientation-dependent relation between angular momentum and magnetic moment. Additionally, materials with a large orbital paramagnetic susceptibility exhibit a strong orbital response to an applied magnetic field and are therefore promising candidates for further studies connecting orbital polarisation with angular-momentum transport. Establishing their potential for orbital-to-spin conversion, orbital pumping, and torque generation will additionally require calculations of transport and dynamical responses. More broadly, our results extend the application of modern, gauge-covariant orbital theory to the gyromagnetic response of metals, strengthening the connection between orbital magnetic responses and orbitronics.

\section*{Methods}
The calculations of susceptibilities and the $g^{\prime}$-tensor were carried out in three computational steps. First, we performed density functional theory (DFT) calculations using the full-potential linearised augmented plane wave (FLAPW) method~\cite{FLAPW}, as implemented in the FLEUR Package~\cite{fleur}.
Exchange–correlation effects were treated within the generalised gradient approximation using the Perdew-Burke-Ernzerhof (PBE) functional~\cite{PBE}. The spin-orbit coupling (SOC) was included self-consistently within the second-variation scheme~\cite{Li1990}, with the spin-quantisation axis chosen along the $z$-axis. We considered bcc V and fcc Pt with lattice constants equal to $3.00 \text{\AA}$ and $3.924~\text{\AA}$, respectively~\cite{go2024}. 
The muffin-tin radius $R_\mathrm{MT}$ was set equal to $2.39~a_{0}$ ($2.29~a_{0}$) where $a_{0}$ denotes the Bohr radius, and the basis functions inside the muffin-tin spheres were expanded in spherical harmonics up to angular momentum $l_\mathrm{max}=10 (12)$ for V (Pt). The plane wave cutoff ($K_\mathrm{max}$) in the interstitial region was set to $5.0~a_{0}^{-1}$ and $3.9~a_{0}^{-1}$ for V and Pt, respectively. The self-consistent non-magnetic electronic structures of V and Pt were obtained using $31\times31\times31$ and $16\times16\times16$ $\mathbf{k}$-point meshes in the Brillouin zone, respectively. The Kohn–Sham and interacting susceptibilities of V and Pt were subsequently evaluated carrying out spin-polarized calculations using $64\times64\times64$ and $32\times32\times32$ $\mathbf{k}$-point grids in the Brillouin zone, respectively.
  
In the second step, maximally localised Wannier functions (MLWFs) were constructed from the converged Kohn–Sham states using the WANNIER90 package~\cite{pizzi2020}.  For V, $30 N_\mathrm{atom}$ states were used to construct $18 N_\mathrm{atom}$ MLWFs while for Pt, $36 N_\mathrm{atom}$ Kohn-Sham states were used to construct $18 N_\mathrm{atom}$ MLWFs. $N_\mathrm{atom}$ denotes the number of atoms in the primitive unit cell.  The initial projections comprised $s$, $p$, and $d$ orbitals. During the disentanglement procedure, frozen energy windows extending up to $7~$eV and $8~$eV above the Fermi level was employed  for V and Pt, respectively. For both metals, the ground-state wavefunctions were generated on a coarse uniform $\mathbf{k}$-point grid of $8 \times 8 \times 8$ and $6 \times 6 \times 6$ for V and Pt respectively, to construct the maximally localised Wannier functions. 
 
In the final step, the susceptibilities and $g^{\prime}$-tensors were calculated using Wannier interpolation, as developed and implemented in the ORBITRANS code~\cite{goOrbitrans}. Brillouin zone integrations were performed on a $250\times 250\times 250$ $\mathbf{k}$-point mesh. 
The calculations were carried using a value of $\Gamma=0.025~$eV 
and Fermi-Dirac distribution for a temperature of 290~K.
In order to suppress spurious contributions arising from near-degenerate band crossings at finite temperature, a small spin Zeeman field of strength $J_z=10^{-4}$~eV was introduced. In addition, contributions for which the energy separation between bands fell below an energy cutoff of $10^{-4}$~eV were discarded to avoid numerical divergences.

\section*{Data availability} 
The data can be made available from the authors upon reasonable request. 

\section*{Code availability}
The ORBITRANS code used in this work will be published in future. Additional workflows are available from the corresponding authors upon reasonable request.

\section*{Acknowledgements} 
The authors would like to thank Insu Baek, Hojun Lee, Hyun-Woo Lee, and David DiVincenzo for fruitful discussions. 
G.A.\ acknowledges the receipt of a grant from the APS-EPS-FECS-ICTP Travel Award Fellowship Programme (ATAP), Trieste, Italy, and from J\"ulich Forschungszentrum, Germany. Y.M.\ acknowledges support by the Deutsche Forschungsgemeinschaft (DFG) in the framework of TRR 288/2 $-$ 422213477 (Project B06), TRR 173/3 $-$ 268565370 (project A11), by the EIC Pathfinder OPEN grant 101129641 ``OBELIX’’, and ORBIS project, grant agreement 101226840, funded by the European Union’s Horizon Europe Programme under the Marie Skłodowska - Curie Actions (MSCA). S.B.\ acknowledges funding from the Deutsche  Forschungsgemeinschaft (DFG) under Germany's Excellence Strategy – Cluster of Excellence Matter and Light for Quantum Computing (ML4Q) EXC 2004/2 – 390534769. We also gratefully acknowledge the J\"ulich Supercomputing Centre for providing computational resources under project jiff40. 
 
\section*{Competing interests}
The authors declare no competing interests.

\section*{Author Contributions}
G.A., M.S., D.G., S.B., Y.M. contributed to the development of the theoretical approach. G.A. performed the calculations. All the authors contributed to the analysis of the results and writing the manuscript.

\section*{Additional Information}
A Supplementary Information file is provided with this manuscript.

\bibliographystyle{naturemag}
\bibliography{bibliography}
\end{document}